\documentstyle[12pt,feynmf,feynmp,openbib,epsfig]{article}

\input epsf
\def\br{\begin{eqnarray}}
\def\er{\end{eqnarray}}
\def\be{\begin{equation}}
\def\ee{\end{equation}}
\def\l{\label}

\def\({\left(}
\def\){\right)}
\def\<{\left\langle}
\def\>{\right\rangle}
\def\lsim{\stackrel{\scriptstyle <}{\phantom{}_{\sim}}}
\def\gsim{\stackrel{\scriptstyle >}{\phantom{}_{\sim}}}
\begin{fmffile}{piond}
\begin{document}

\vspace{1cm}
\begin{center}
{\large\bf Medium effects in the  pion-pole mechanism
($\gamma\gamma \rightarrow \pi^0 \rightarrow \nu_R\bar{\nu}_L
(\nu_L \bar{\nu}_R)$) of neutron
star cooling}\\[.4cm]
F. Arretche,  A. A. Natale \\
Instituto de F\'{\i}sica Te\'orica, Universidade Estadual
Paulista,
Rua Pamplona, 145, 01405, S\~ao Paulo, SP, Brazil \\
and D. N. Voskresensky \\

Moscow Institute for Physics and Engineering, Russia, 115409
Moscow
\\
\end{center}
\thispagestyle{empty} \vspace{1cm}

\begin{abstract}
Nuclear medium effects in the neutrino cooling of neutron stars
through the  reaction channel $\gamma\gamma \rightarrow \pi^0
\rightarrow \nu_R\bar{\nu}_L (\nu_L \bar{\nu}_R)$ are
incorporated. Throughout the paper we discuss different
possibilities of right-handed neutrinos,  massive left-handed
neutrinos and standard massless  left-handed neutrinos (reaction
is then allowed only with medium modified vertices). It is
demonstrated that multi-particle effects suppress the rate of this
reaction channel in the dense hadron matter by $6 - 7$ orders of
magnitude that does not allow to decrease existing experimental
upper limit on the corresponding $\pi^0 \nu\bar{\nu}$ coupling.
Other possibilities of the manifestation of the given reaction
channel in different physical situations, e.g. in the quark color
superconducting cores of the most massive neutron stars, are also
discussed. We demonstrate that in the color-flavor-locked
superconducting phase for temperatures $T\lsim (0.1\div 10)$~MeV
(depending on the effective pion mass and the decay width)  the
process is feasibly the most efficient neutrino cooling process,
although the absolute value of the reaction rate is rather small.

\end{abstract}

\newpage

\section{Introduction}

\par Many years ago Pontecorvo and Chiu and Morrison\cite{ponte}
suggested that the process $\gamma \gamma \rightarrow \nu
\bar{\nu}$ might play an important role as a mechanism for stellar
cooling. Gell-Mann\cite{gm} subsequently showed that this process
is forbidden in a local (V-A) theory. However, it can occur at the
one-loop level which has been computed by Levine\cite{levine} for
an intermediate-boson (V-A) theory, and the stellar energy loss
rate through $\gamma \gamma \rightarrow \nu \bar{\nu}$ was found
to be smaller than the rates for competing processes (pair
annihilation $e^+ e^- \rightarrow \nu \bar{\nu}$ and
photo-neutrino production $\gamma e \rightarrow e \nu \bar{\nu}$).
This result is not modified when the cross section of the above
process is computed in the standard model, as was shown by
Dicus\cite{dicus}. Only for very peculiar neutrino coupling to
photons or unnaturally large neutrino masses this reaction
overwhelms the result of the standard model\cite{natale}.

There is still another possibility proposed by Fischbach {\it et
al.} \cite{fisch}, where the reaction $\gamma \gamma \rightarrow
\nu \bar{\nu}$ could be significant. This is the case when the
process is mediated by a pseudoscalar resonance and the latter
decays into $\nu \bar{\nu}$ due to the existence of right-handed
neutrinos or due to new interactions beyond the standard model. It
was assumed that in astrophysical conditions only the pion
resonance could be important (next in mass not strange
$\eta$-resonance is too heavy in standard conditions) and the
process was termed the pion-pole mechanism. Thus the process which
we will continue to  discuss in this paper  is $\gamma \gamma
\rightarrow \pi^0 \rightarrow \nu \bar{\nu}$.

Of course, if the temperature is high enough, and on the other
hand, the pion dispersion relation in matter allows for the
quasiparticle spectrum branch, there appears a significant number
of  thermally equilibrated pion quasiparticles. Then the process
$\pi^0\rightarrow \nu \bar{\nu}$ may also be important. In this
process the initial thermally equilibrated pion is on its
mass-shell modified in the matter. In the process $\gamma \gamma
\rightarrow \pi^0 \rightarrow \nu \bar{\nu}$ the initial reaction
states contain no pion, the virtual pion only transfers the
interaction from thermally equilibrated photons in the
$\gamma\gamma$ annihilation process to produced $\nu\bar\nu$. As
we will see below, the process $\gamma \gamma \rightarrow \pi^0
\rightarrow \nu \bar{\nu}$ has an output of the energy $Q$ varying
with the temperature as $Q\propto T^n$ where the power $n$ changes
with the temperature typically from  $n=3$ for rather high
temperature ($T$ is still much smaller than the pion mass
$m_{\pi}=140~$MeV) to $n=11$ for low temperatures. The process
$\pi^0\rightarrow \nu \bar{\nu}$ having essentially larger phase
space volume (one particle in the initial state) yields however
the exponentially suppressed output of the energy, $Q\propto
T^{3/2} e^{-m_{\pi}/T}$ at $T<m_{\pi}$  since the initial particle
is massive, in difference with photons. Concentrating on the
discussion of the rate of the $\gamma \gamma \rightarrow \pi^0
\rightarrow \nu \bar{\nu}$ reaction channel we shall also compare
it with the rate of the competing $\pi^0\rightarrow \nu \bar{\nu}$
process and other relevant processes.

Temperatures of the order of $10 - 60$ MeV are expected in
interiors  of  proto-neutron stars formed in supernova (SN)
explosions during the early cooling phase of the proto-neutron
star evolution, according to existing numerical
simulations\cite{burrows}. It is exactly in such a situation the
pion-pole mechanism is the most effective ($T\sim m_{\pi}/a $,
$a\sim 2\div 50$). Indeed, in Ref.\cite{sn} the mechanism was
applied to the case of supernova SN1987A and it was found to be a
quite important process even if the pion partial production rate
of neutrinos ($\Gamma (\pi^0 \rightarrow \nu_R \bar{\nu}_L (\nu_L
\bar{\nu}_R $ ))) is many orders of magnitude smaller than the
presently accepted value of the experimental upper limit. In case
of $\nu_L \bar{\nu}_R $ with the vacuum vertex  the process rate
is proportional to the squared neutrino mass. The result \cite{sn}
was criticized by Raffelt and Seckel \cite{raffelt} basing on the
fact that in the calculation of Ref.\cite{sn} it was used the
vacuum value for the total pion width, $\Gamma_{\pi}^{vac}$, which
is a tiny quantity, $\Gamma_{\pi}^{vac}\simeq 0.58\cdot
10^{-7}m_{\pi}$. They argued that in supernova cores pion states
are damped mostly by nucleon absorption rather than by the free
decay so that the width in the medium ($\Gamma_{\pi}^{med}$) is
much larger than in vacuum and the pion-pole mechanism should be
strongly suppressed. Besides, in the medium the process may go on
massless left-handed neutrinos  through the intermediate nucleon
particle $-$ nucleon hole states, i.e. as  $\gamma\gamma
\rightarrow \pi^0\rightarrow nn^{-1}\rightarrow \nu_L \bar{\nu}_R$
reaction channel. The processes  $\pi^0_{virt} \rightarrow
nn^{-1}\rightarrow \nu_L \bar{\nu}_R$  on virtual pions have been
discussed in \cite{VS86}.

Without any doubt the neutrino emission from the dense hadronic
component in neutron stars is the subject of strong modifications
due to collective effects in the nuclear matter \cite{dv}, and it
is interesting to quantitatively know how much the pion-pole
mechanism is influenced by the properties of the dense medium and
even, if the process is strongly suppressed, what is the main
effect causing this suppression. In this work we compute the
cooling rate due to the neutrino emission through the pion-pole
mechanism in the reaction $\gamma\gamma\rightarrow\pi^0
\rightarrow \nu_R \bar{\nu}_L $ (or $\nu_L \bar{\nu}_R )$ for the
conditions of proto-neutron stars including the effects of dense
medium in the pion polarization operator (Section 2). Then  we
discuss other relevant reaction channels. In order to clarify
their connection to the process $\gamma\gamma\rightarrow\pi^0
\rightarrow \nu_R \bar{\nu}_L $ (or $\nu_L \bar{\nu}_R )$ we
review in the Appendix the optical theorem formalism, see the
discussion in \cite{dv}, that allows to calculate consistently the
reaction rates including complicated medium effects. Continuing to
study possible physical situations where the pion pole cooling
mechanism could be important we consider in Section 3 a possible
consequence of the pion pole mechanism in the case of neutron
stars with the color superconducting quark cores. We argue that in
the color-flavor-locked superconducting phase  for temperatures
$T\lsim (0.1\div 10)$~MeV the process is feasibly the most
efficient neutrino cooling process, although the absolute value of
the reaction rate is rather small. Then we draw our conclusion.

\section{Emissivity of the process $\gamma\gamma\rightarrow\pi^0
\rightarrow \nu\bar{\nu}$ from the hadronic matter}

\par The cross section of the process $\gamma \gamma \rightarrow \pi^0
\rightarrow \nu_R \bar{\nu}_L (\nu_L \bar{\nu}_R )$ in vacuum is
given by
\be \sigma_{\pi}^{vac} (s) = \frac{8\pi (s^2/m_{\pi}^4) \Gamma
(\pi^0 \rightarrow \gamma \gamma)\Gamma (\pi^0 \rightarrow \nu
\bar{\nu}) F(s)}{(s-m_{\pi}^2)^2 + (m_{\pi} \Gamma_{\pi}^{vac})^2
}, \l{crosssec} \ee
where $F(s)$ is an unknown function of $s \equiv$(c.m.
energy)$^2$, representing the product of the vertex function for
the off-shell processes $\gamma \gamma \rightarrow \pi^0$ and
$\pi^0 \rightarrow \nu \bar{\nu}$, constrained to
$F(s=m_{\pi}^2)=1$. Following Ref.\cite{fisch} we may assume
$F(s)\simeq 1$ off mass-shell, that seems to be a reasonable
approximation for the energies and momenta that we are dealing
with. $\Gamma (\pi^0 \rightarrow \gamma \gamma)$ is the partial
width of the pion decay into two photons. Here we assume that
$\Gamma (\pi^0 \rightarrow \gamma \gamma)$ is approximately equal
to the total pion width ($\Gamma_{\pi}^{vac}$), for which we use
the experimental value.  The pion partial width into neutrinos
$\Gamma (\pi^0 \rightarrow \nu \bar{\nu})$ has the following
experimental upper limit $\Gamma (\pi^0 \rightarrow \nu
 \bar{\nu})/\Gamma_{\pi}^{vac} \,<
\, 8.3 \times 10^{-7}$ \cite{pdg}. In (\ref{crosssec}) one
recognizes the free pion propagator modulus squared $\mid
D^{vac}_{\pi}\mid^2$ entering squared matrix element of the
reaction under consideration.

Strictly speaking, in application to the nuclear medium   all the
terms in (\ref{crosssec}) should be modified. E.g., the partial
width $\Gamma (\pi^0 \rightarrow \gamma\gamma)$ may increase with
the temperature \cite{CL}. The pion partial width into neutrinos
$\Gamma (\pi^0 \rightarrow \nu \bar{\nu})$ is determined by the
corresponding $\pi^0 - \nu\bar\nu$ coupling. The squared vertex is
averaged over the neutrino--antineutrino phase-space volume, see
Appendix. If we knew the explicit form of the coupling we could
explicitly calculate $\Gamma (\pi^0 \rightarrow \nu \bar{\nu})$
and its temperature dependence. However the main modification
comes from the change of the pion propagator in dense nuclear
medium due to the pion pole. Therefore, below we consider only
this modification.  Then, one should replace in (\ref{crosssec})
the free pion propagator  $ D^{vac}_{\pi}$ by the in-medium one
\be D_{\pi^0}^R = \frac{1}{[ \omega^2 - m_{\pi}^2 - k^2 -
\Pi_{\pi^0}^R (w,k,\rho , Y , T)]}, \l{piprop} \ee
where  $\Pi_{\pi^0}^R (w,k,\rho , Y , T)$ is the total retarded
polarization operator of the neutral pion, dependent on  the pion
energy $\omega$, momentum $k$, baryon (nucleon in our case)
density $\rho$, isotopic composition $Y=Z/(N+Z)$, and the
temperature $T$.

If  photons are in thermal equilibrium, the energy loss rate of
the process $\gamma \gamma \rightarrow \pi^0 \rightarrow \nu_R
\bar{\nu}_L (\nu_L \bar{\nu}_R )$ occurring in dense interior of
the proto-neutron star, at which the energy is converted into
neutrino pairs, is given by
\be Q^{med} = \frac{4}{(2\pi)^6} \int \frac{d^3 \,
k_1}{[exp(\omega_1/k_B T) - 1]} \frac{d^3 \,
k_2}{[exp(\omega_2/k_B T) - 1]} ( \omega_1 + \omega_2) {\it
v}_{rel} \sigma_{\pi}^{med} , \l{qrate} \ee
where $\omega_1$, $\omega_2$ are the photon energies, $\vec{k}_1$
and $\vec{k}_2$ are their momenta, $\theta$ is the angle between
photons,  ${\it v}_{rel}$ is the relative-velocity factor
\be {\it v}_{rel} = |(\vec{k}_1 \cdot \vec{k}_2 - \omega_1
\omega_2)/ \omega_1 \omega_2 | = 1 - \cos \theta , \l{vel} \ee
and $\sigma_{\pi}^{med}$ is the medium dependent cross section of
the process given by (\ref{crosssec}) with $D_\pi^{vac}$ replaced
by $D_{\pi^0}^R$. Eq. (\ref{qrate}) is the straightforward
generalization of the result  \cite{fisch,sn} obtained with the
help of the replacement $\sigma_{\pi}^{vac} \rightarrow
\sigma_{\pi}^{med}$.

The energy loss rate (\ref{qrate})  can be presented in the
following form
\br Q^{med} \simeq 4.21 \times 10^{22}
%\(\frac{16}{\pi^3}\)
\(\frac{\Gamma (\pi^0 \rightarrow \gamma \gamma)}{m_{\pi}}\)
\(\frac{\Gamma (\pi^0 \rightarrow \nu \bar{\nu})}{m_{\pi}}\)
%\(\frac{1 \, GeV^2}{m_{\pi}^2}\)\\
%\nonumber && \times
%\( \frac{k_B T}{m_{\pi}}  \)^4  (k_B T)^7
T_9^{11}I(\tau ),\,\,\,  erg/(cm^3 \cdot s),  \l{qpi} \er
$m_{\pi}=140$~MeV, $T_9 =T/10^{9}K$,
% The dimensional $k_B T$ term enters in $erg$,
$\tau = (k_B T/m_{\pi})$ and $I(\tau )$ is given by the
dimensionless integral
\br\label{integ} I(\tau ) &=& \int_0^\infty \frac{x_1^4 \,
dx_1}{e^{x_1} - 1}\int_0^\infty \frac{x_2^4 \, dx_2}{e^{x_2} - 1}
\\ \nonumber && \times \int_\kappa^2 \frac{du(x_1+x_2)
u^3}{(2x_1x_2u \tau^2 - 1 - m_{\pi}^{-2}Re\Pi^R_{\pi^0})^2
+(m_{\pi}^{-1}\Gamma_\pi )^2 },
%\label{integ}
\er
where $\kappa =\frac{2}{x_1 x_2}(m_\nu/k_B T)^2$, $m_\nu$ is the
neutrino mass, $x_{1,2} = \omega_{1,2}/k_B T$, $u=1-\cos \theta$,
$m_{\pi}^{-1}\Gamma_\pi =m_{\pi}^{-1}\Gamma_\pi^{vac} -
m_{\pi}^{-2} Im\Pi^R_{\pi^0}$. Further we assume that $m_\nu/k_B T
\ll 1$ and then put the lower limit ($\kappa$) of the integral in
the variable $u$ equal to zero, that corresponds to using
$m_\nu=0$ in all the phase space calculations. However we take
$m_\nu \neq 0$ into account evaluating the $\Gamma (\pi^0
\rightarrow \nu_L \bar{\nu}_R)$ width. For the further convenience
we assume that $Re\Pi^R_{\pi^0}$ and $Im\Pi^R_{\pi^0}$ are the
real and imaginary parts of $\Pi_{\pi^0}^R (w,k,\rho , Y , T)$
describing only the strong interaction processes. Therefore we
separated in (\ref{integ}) the value $\Gamma_{\pi}^{vac}$ which is
due to the electromagnetic and the weak interaction.
%, and $\gamma_\pi^{vac} = m_{\pi}\Gamma_{\pi}^{vac}$.
With $I(\tau )$ replaced to $I^{vac}(\tau )$ (putting
$Re\Pi^R_{\pi^0}$ and $Im\Pi^R_{\pi^0}$ to zero) we reproduce the
results of Refs.\cite{fisch,sn}. One easily finds the
corresponding asymptotic expression \br\label{asym} I\sim
I^{vac}(\tau\ll 0.1)\simeq 23040 \,\zeta (6)\zeta (5)\left[
1+96\tau^2 \zeta (7)/\zeta (5)\right]\,,\er $\zeta$ is the Riemann
function, $\zeta (5) \simeq 1.037$,  $\zeta (6)\simeq 1.017$,
$\zeta (7)\simeq 1.008$. In order to get (\ref{asym}) we dropped
$\Gamma_\pi$ and $Re\Pi^R$ and expanded (\ref{integ}) in the value
$2x_1 x_2 u\tau^2 \ll 1$. Since typical values $x_1\sim x_2\sim 6$
in the resulting integral, the limit expression is valid for $\tau
\ll 0.1$. As we checked numerically the limit (\ref{asym}) is
actually achieved only at $\tau \lsim 0.01$, if $\Gamma_\pi$ is as
small as $\Gamma^{vac}_\pi$. Using Eq. (\ref{asym}) and the
evaluation of $\Gamma (\pi^0 \rightarrow \gamma \gamma)$ we obtain
\be\label{vacq} Q^{vac}(\tau \lsim 0.01)\simeq 0.6\cdot
10^{20}\left(m_{\pi}^{-1}\Gamma( {\pi}^0 \rightarrow
\nu\bar{\nu})\right)T_9^{11},\,\,erg/(cm^3 \cdot s) \ee at small
temperatures.

When the temperature increases  the denominator becomes to be near
the pole. Then dividing and multiplying $I(\tau)$ by
$\Gamma_{\pi}$ and using the corresponding presentation of the
$\delta$-function we roughly estimate \br\label{asym1}
I\left((\frac{m_{\pi}}{10\Gamma_{\pi}})^{1/4}\gsim\tau \gsim
a^{-1} \right)\sim \frac{0.6(e^{-1/(2\tau)}|\mbox{ln} (2\tau
)|+1)e^{-1/(2\tau)}}{\tau^8 m_{\pi}^{-1}\Gamma_{\pi}} ,\,\,\,
a\sim 10\div 10^2 . \er From the latter estimate we recognize the
resonant character of the rate. $I^{vac} (\tau \sim 0.1)$ is $\sim
10^9$ times larger compared with  $I^{vac} (\tau =0)$ for
$\Gamma_{\pi}\sim \Gamma_{\pi}^{vac}$.

For higher temperatures assuming $2x_1 x_2 u \tau^2 \gg 1$ we
estimate \br\label{asym11} I\left(\tau \gg
(\frac{m_{\pi}}{10\Gamma_{\pi}})^{1/4}\right)\simeq 12 \zeta
(3)\zeta (4) /\tau^4 ,\,\,\,\, \zeta (3) \simeq 1.202, \,\,\,
\zeta (4)\simeq 1.082 ,\er that produces $ Q^{vac}\simeq
Q^{med}\simeq 2.6\cdot 10^{29}\left(m_{\pi}^{-1}\Gamma( {\pi}^0
\rightarrow \nu\bar{\nu})\right)T_9^{7},\,\,erg/(cm^3 \cdot s).$

In the so called "standard scenario" of the neutron star cooling
one considers the modified Urca process as the most efficient
process. The emissivity of the modified Urca process is estimated
\cite{FM} with the help of the free pion propagator as $Q^{MU}
\sim 10^{21}T_9^{8}$ $erg/(cm^3 \cdot s).$  Comparing it with
(\ref{qpi}), (\ref{asym1}) we see that  for $T\gsim 10$ MeV the
process under consideration would be much more efficient process
than the modified Urca process, if the pion width and the mass in
medium were not essentially changed compared to the vacuum values.
However in reality the width $\Gamma^{vac}_{\pi}$ is replaced to a
much larger medium value. Also  the pion mass is modified  by the
polarization effect (thereby the pion pole begins to manifest
itself at $T\gsim m_{\pi}^{eff}/a$, rather than at $T\gsim
m_{\pi}/a$). Thus one may expect that with taking into account of
the medium effects the estimation (\ref{asym}) is not essentially
modified whereas the value (\ref{asym1}) must be significantly
suppressed mainly due to the suppression of the width.

Techniques for the description of collective effects in dense
hadronic matter have been developed in the last
decades\cite{migdal1,migdal2}. E.g., the medium effects appearing
in the pion propagator have been discussed in detail in
Ref.\cite{dvnp} (see Appendix B of that work). For the
temperatures and pion energies with which we are concerned the
main modification of the pion polarization operator is due to the
density dependence rather than the temperature dependence and it
is, thereby, a reasonable approximation to put $T=0$ in
$\Pi^R_{\pi^0} (\omega,k,\rho ,Y, T)$. Depending on the values of
the typical pion energy and momentum different terms can be
important in the pion polarization operator. At small pion
energies $\omega /m_{\pi}\ll 1$ and for typical momenta $k \sim
p_{FN}$,  $p_{FN}$ is the Fermi momentum of the nucleon, the
nucleon particle - hole contribution is attractive and the
dominant one, what results in the softening of the virtual pion
mode with increase of the density and leads to the possibility of
the pion condensation at $\rho
>\rho_c > \rho_0$, cf. \cite{migdal1,migdal2,dvnp}. In our case $\omega >k$,
as follows from the reaction kinematics, and the nucleon particle
- hole contribution is minor. Then the main terms in the pion
polarization operator are  the $\Delta$-particle - nucleon hole
part and the regular part related to  more complicated
intermediate states. Thus, $\Pi^R_{\pi^0}\simeq \Pi^R_{\Delta}
+\Pi^R_{reg},$ and the partial contributions are given by
\cite{dvnp}:
\be Re \, \Pi^R_{\Delta} (\omega,k,\rho , Y, T=0)\simeq -
\frac{B_0 \Gamma(g'_{\Delta})}{\tilde{\omega}^2_\Delta (t) -
\omega^2}, \l{repiad} , \ee
in units $\hbar =c =1$, $B_0 \simeq 2.0~m_{\pi}(\rho /\rho_0)
\Gamma^2_{\pi N \Delta} k^2 \tilde{\omega}_\Delta (t)$, $\rho$ is
expressed in units $\rho_0$, $\rho_0 \simeq 0.5~m_{\pi}^3$ is the
saturation nuclear density, the form factor $\Gamma^2_{\pi N
\Delta} \approx \Gamma^2_{\pi N N} / \beta \simeq 1/\beta$ for the
rather small momenta of our interest, $\beta \simeq 1+0.23 k^2
/m_{\pi}^2$ is an empirical factor taking into account a
contribution of the high-lying nucleon resonances. The
nucleon-$\Delta$ isobar  correlation factor $\Gamma(g'_{\Delta} )$
is given by
\be \Gamma(g'_{\Delta}) \simeq \left[ 1 +
\frac{C}{\tilde{\omega}^2_\Delta (t) - \omega^2} \right]^{-1} ,
\l{ggl} \ee
with $C\simeq 0.9~m_{\pi}^2 (\rho /\rho_0 ) \Gamma^2_{\pi N
\Delta}$ and
\be \tilde{\omega}_\Delta (t) \simeq 2.1~m_{\pi} \left( 1 +
\frac{2.1~m_{\pi}}{2m^{\star}_N}\right) + \frac{t
}{2m^{\star}_N},\,\,\,\,t= k^2 -\omega^2 , \l{omed} \ee
$m^{\star}_N (\rho)$ is the effective mass of the nucleon
quasiparticle,
\be Im \, \Pi^R_{\Delta} (\omega,k,\rho , Y,T=0) \simeq -
\frac{2\omega B_0 \Gamma^2(g'_{\Delta}) \gamma_0 k^3}
 {[\tilde{\omega}^2_\Delta (t) - \omega^2]^2},
\l{impiad} \ee
$\gamma_0 k^3$ takes into account  the $\Delta$-isobar width, with
an empirical value $\gamma_0 \simeq 0.08~m_{\pi}^{-2}
\Gamma^2_{\pi N \Delta}$. In the numerical evaluations below we
for simplicity assume $\tilde{\omega}^2_\Delta (t) \gg \omega^2$
that is a reasonable approximation for the conditions we are
dealing with.

The regular part of the pion polarization operator is yet more
model dependent. Its value is recovered with the help of the
pionic atom data, cf. \cite{BFG},  and a procedure of going off
mass-shell. We present it in the following form, cf. \cite{dvnp},
\be Re \, \Pi_{reg}^R (\omega,k,\rho , Y,T=0)\simeq \left(
-0.25~\omega^2 + 0.25~m_{\pi}^2 + 0.5~ k^2
\right)\frac{\rho}{\rho_0} , \l{reregpi} \ee
\br && Im \, \Pi_{reg}^R (\omega,k,\rho , Y,T=0) \simeq
-0.15~m_{\pi}^2 \left(\frac{\rho}{\rho_0}\right)^2 \xi \nonumber \\
&&-\frac{0.09(\rho/\rho_0 )^2 k^2 \xi
%%\Gamma^{2\alpha}_{\pi N N}
} {[1+ 0.38 (\rho/\rho_0 ) - 0.047 (\rho/\rho_0 )^2 \xi)]^2}.
\l{imregpi} \er
The factor $\xi$ which we inserted in these expressions compared
to those of \cite{dvnp} takes into account asymmetry of the
isotopic composition of the proto-neutron star matter. Near the
pion mass-shell, with $\xi =4Y(1-Y)$ we approximately describe
results given by the potentials I-III used in \cite{BFG} for the
pion atoms and with $\xi =1$, the results for the potential IV.
For the value $Y\simeq 0.4$ typical for initial stage of
proto-neutron star cooling in both mentioned cases one can put
$\xi \simeq 1$. The real and imaginary terms shown above are the
ones which enter Eq.(\ref{integ}), being responsible for the
density effects, where we put $\xi \simeq 1$. We also used that
$\Pi^R_{\pi^0} (\omega,k,\rho , Y,T)=
\frac{1}{2}\left[\Pi^R_{\pi^-} (\omega,k,\rho , Y,T)+\Pi^R_{\pi^+}
(-\omega,-k,\rho , Y,T)\right]$, and the linear in $\omega$ terms
entering $Re\Pi_{\pi^\pm}^R$ for $Y\neq 1/2$ do not contribute to
$Re\Pi^R_{\pi^0}$.

\begin{figure}[ht]
\begin{center}
\includegraphics{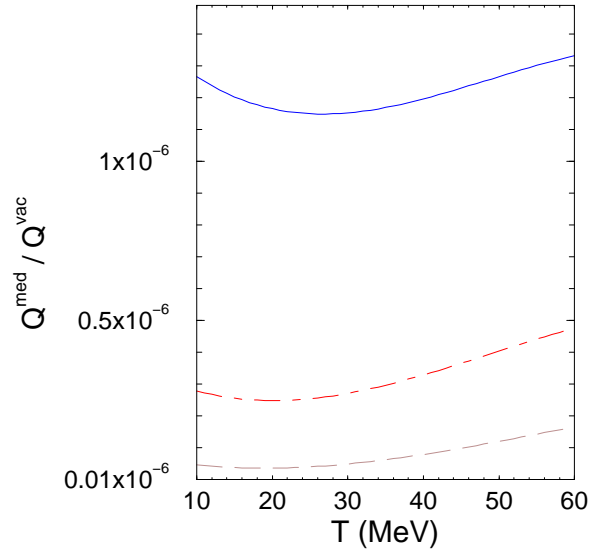}
\caption[dummy0]{Ratio of energy loss $Q^{med}/Q^{vac}$ as
function of temperature for $\rho = 0.5\rho_0 $ (solid line),
$\rho = \rho_0 $ (dot-dashed line) and  $\rho = 2\rho_0 $ (long
dashed line). } \label{fig1}
\end{center}
\end{figure}

We computed the energy loss rate including  medium effects in the
pion polarization operator. In Figure \ref{fig1} we present the
results for the ratio of   the energy loss rate ($Q^{med}$) due to
the pion-pole mechanism calculated with medium effects taken into
account in the pion polarization operator  to the one ($Q^{vac}$)
computed with $Re \, \Pi^R =0$ and the vacuum pion decay width
(cf. (\ref{vacq})),  for three values of the nuclear matter
density: $\rho = (0.5; 1; 2) \rho_0 $. We used $m^{\star}_N \simeq
(0.9; 0.85; 0.7) m_N$ for those densities and  put $m_{\pi} \simeq
140$~MeV.
%%Here and in the following we refer to $Q^{vac}$ as the emissivity computed
%%with the vacuum value for the pion width and  $m_{\pi} \simeq 140$~MeV.
%
The ratio $Q^{med}/Q^{vac}$ is depicted in Figure \ref{fig1} as a
function of temperature. This ratio does not depend on an unknown
value of $\Gamma (\pi^0 \rightarrow \nu\bar{\nu})$. We see that
the nuclear matter effects decrease the output of the energy
typically by six to seven orders of magnitude depending on the
temperature and the density, in agreement with the expectations of
Raffelt and Seckel~\cite{raffelt}. We also show in Table
\ref{tab1} the results for the dimensionless integral of
Eq.(\ref{integ}).

\begin{table}

\begin{center}

\begin{tabular}{|l|c|c|c|c|c|c|c|c|c|c|}
%                     &\multicolumn{10}{|c|}{$\tau$}     \\
\hline
$T(MeV)$ & 10 & 20 & 30 & 40 & 50 & 60 \\
\hline
$\rho=0.5\rho_{0}$ & 468277 & 605757 & 157430 & 41569 & 12907 & 4713 \\
$\rho=\rho_{0}$    & 115894 & 131020 & 38552 & 11663 & 4149  & 1722 \\
$\rho=2\rho_{0}$  & 25594 & 22432 & 8165 & 3044 & 1293 & 621 \\
\hline
\end{tabular}
\end{center}

\caption{Numerical results for the dimensionless integral
$I(\tau)$ of Eq.(\ref{integ}).} \label{tab1}
\end{table}

Qualitatively,  the reasons for the enormous decrease of the
energy loss rate are related to the strong pion absorption in
nuclear matter, as has been  predicted in Ref.\cite{raffelt}. The
total pion width at appropriate pion energies and momenta grows up
to tens of MeV with the density, that is orders of magnitude
larger than the vacuum contribution $\Gamma_\pi^{vac}$. Other
reason is the energy-momentum dependence of the real part of the
pion polarization operator. This is also not  a small change
because the mechanism is totally dependent on the resonant
behavior, and outside the resonance the contribution sharply
decreases. We do not need to be on the top of the resonance,
however we cannot be too far away either. Medium effects push the
typical pion energy to a larger value ($Re\Pi^R (\omega , k)
>0$ at pion energies and momenta in the region of the pole).
%% and there is no temperature
%%in the proto-neutron star high enough to keep the process near the
%%resonance.

The ratio $\Gamma (\pi^0 \rightarrow \nu \bar{\nu})/ \Gamma (\pi^0
\rightarrow\gamma \gamma)$ in many models of elementary particles
beyond the standard one is proportional to the neutrino mass
($m_{\nu}$). In Ref.\cite{sn} a limit on the neutrino mass was
obtained due to the strong constraint on  $\Gamma (\pi^0
\rightarrow \nu \bar{\nu})$. However, with medium effects included
into consideration no astrophysical limit on $m_{\nu}$ better than
the existent ones can be obtained. A nonzero neutrino mass induces
other mechanisms much more efficient than the one generated by the
pion resonance in this situation, and these mechanisms provide
then tighter constraints on the neutrino masses (see, for
instance, Ref.\cite{revraf}).

Above we discussed only the role of the reaction channel
$\gamma\gamma \rightarrow \pi^0 \rightarrow \nu \bar{\nu}$. But
there are other competing reaction channels. The relation between
all these processes becomes to be clear if one applies the so
called optical theorem formalism for the calculation of the
reaction rates, see discussion in the Appendix.

First, if the pion can be described within the quasiparticle
approximation in some region of its energy and momentum, one has a
contribution of the process $\pi^0\rightarrow \nu\bar\nu$ whose
rate is given by,
%%\begin{equation}\label{qboltz}
%% Q^{\pi^0\nu\bar\nu} \sim  m_{\pi}^{*\,2}
%% \left( \frac{m_{\pi}^* k_B T}{2\pi \alpha} \right)^{3/2}
%% e^{-m_{\pi}^* /(k_B T)}    \Gamma (\pi^0 \rightarrow \nu
%%\bar{\nu})/m_{\pi}.
 %%\end{equation}
%
\br\label{pin} Q^{\pi^0\nu\bar\nu}\sim \left( \frac{\Gamma
(\pi^0\rightarrow \nu\bar\nu )}{m_{\pi}} \right)
n_{\pi}m_{\pi}^{*2} , \er
$$n_{\pi}=\frac{1}{v_{\pi}^3}\left(\frac{
m^*_{\pi}k_B T}{2\pi}\right)^{3/2}e^{-m^*_{\pi}/(k_B T)}.$$ Here
we used the dispersion relation $\omega^2 =(m_{\pi}^* )^2
 +v_{\pi}^2 k^2$ for small momenta $k\sim k_B T$ typical in this
 reaction. The effective mass $m_{\pi}^* \sim m_{\pi}$ and the velocity
 $v_{\pi} <1$ are calculated
 according to eqs. (\ref{repiad}) -- (\ref{reregpi}).
The effects of the partial $\Gamma (\pi^0 \rightarrow \gamma
\gamma)$ width are not present in this process.

Second, the presence of finite width of the virtual pion $Im
\Pi_{\pi^0}\neq 0$ due to the strong interaction means also a
finite contribution of the reaction $\pi^0_{virt} \rightarrow \nu
\bar{\nu}$ that does not need  $\gamma\gamma$ and pion
quasiparticle  states, but relates to the corresponding nucleon
states, cf. \cite{VS86,dv} (it is clearly seen after the cut of
the in-medium pion Green function describing propagation of the
in-medium pion). Presence of the imaginary part of the
nucleon-hole term of the pion propagator (with the full vertex) in
appropriate energy-momentum region would lead to a contribution of
the processes  $N\rightarrow N\nu\bar{\nu}$, $NN\rightarrow
NN\nu\bar{\nu}$ going via the virtual pion, the later couples $N$
with $\nu\bar\nu$. The presence of the imaginary part of the
regular term in the pion polarization operator corresponds to more
involved multi-nucleon and multi-pion states.

Third, the processes as $\gamma\gamma \rightarrow \pi^0
\rightarrow nn^{-1} \rightarrow \nu_L\bar{\nu}_R$ ($n^{-1}$ is the
neutron hole) with ordinary left-handed massless neutrinos are
possible leading to substantially larger contribution to the
emissivity than that related to the massive left-handed neutrinos
in the reaction $\gamma\gamma \rightarrow \pi^0 \rightarrow \nu_L
\bar{\nu}_R$. The reaction channel $\gamma\gamma \rightarrow \pi^0
\rightarrow nn^{-1} \rightarrow \nu_L\bar{\nu}_R$ may also lead to
substantially larger contribution to the emissivity than that
related to the right-handed neutrinos (depending on the value of
the $\Gamma (\pi^0 \rightarrow \nu_R \bar{\nu}_L )$)  in the
reaction $\gamma\gamma \rightarrow \pi^0 \rightarrow \nu_R
\bar{\nu}_L$ which we have considered. Besides, the process $\pi^0
\rightarrow nn^{-1} \rightarrow \nu_L\bar{\nu}_R$ going on the
left-handed massless neutrinos is possible going on the thermal
pion quasiparticle. The latter rate can be calculated in complete
analogy to that computed for the massive pseudo-Goldstone (photon)
mode in \cite{VKK98}. The processes involving pion but going via
$nn^{-1} - \nu\bar{\nu}$ coupling (with usual $V-A$ coupling of
$N$ to $\nu_L\bar{\nu}_R$) exist only due to the medium
modification of the vertices. Another possibility, which will be
discussed below, is that due to the breaking of the Lorentz
invariance in matter there might appear two pion decay coupling
constants, the temporal  one, $f_T$, and the space one,
$f_S$,\cite{PR,RS}. Their finite difference would also lead to a
contribution  to the $\pi^0$ coupling with the left-handed
neutrinos. Although all these relevant processes are related to
each other, they are characterized by quite different phase-space
volumes, kinematics and/or vertices. Having selected the
$\gamma\gamma \rightarrow \pi^0 \rightarrow \nu\bar\nu$ channel
among others, we considered the enhancement of the rate due to the
particular kinematics of this process (massless particles in
initial states). But, as we have shown, the presence of not as
small  pion width in the hadron matter significantly suppresses
the rate.

Concluding this section we once again stress that  although the
temperature dependence of the
 energy output in the process $\gamma\gamma \rightarrow
\pi^0\rightarrow \nu\bar\nu$ is quite different from those for the
other processes, where also enters the $\pi\nu\bar\nu$ coupling,
the absolute value of the rate is proved to be strongly suppressed
due to the presence of not as small pion width in the dense and
heated nucleon matter. Thus, due to the width effects, the process
has a minor role in the cooling of the dense hadron matter.

\section{Emissivity of the process $\gamma\gamma\rightarrow\pi^0
\rightarrow \nu\bar{\nu}$ from the color-flavor-locked phase}

\par There is  another interesting possibility in connection with the
process under consideration. As has been recently shown, the
interiors of the most massive neutron stars may contain  dense
quark cores which are high temperature color superconductors with
the critical temperature $T_c \simeq 0.6 \Delta_q \lsim 50 $~MeV,
where $\Delta_q$ is the pairing gap between colored quarks, see
the review papers \cite{NS,NSa}. Also the possibility of
self-bounded strange quark stars, being the diquark condensates,
is not excluded, see \cite{NS1}. The diquark condensates may exist
in different phases. The most symmetric phase of dense quark
matter is  the so called the color-flavor-locked  phase. This
phase becomes to be energetically preferable in the large density
limit. The neutrino processes are significantly suppressed in this
phase due to presence of the large diquark gap and absence of the
electrons \cite{BKV,BGV}. On the other hand, it was shown, cf.
\cite{SS99,NSa}, that this phase contains low-lying excitations
with the pion, kaon and $\eta$, $\eta^{\prime}$ quantum numbers
($m^*_{\pi^0 }, m^*_{K^0}, m^*_{\bar{K}^0}, m^*_{\eta},
m^*_{\eta^{\prime}}$ are in the range $(1 - 100)$~MeV,  all
in-medium masses are indicated by $m^*$). Excitation spectra, as
they are calculated in the mentioned works, contain no widths
effects, the $i$-meson Green function is given by $D_i
=\frac{1}{(\omega^2 -k^2 /3 -m_i^{*2})}$. However, in any case
there is a width contribution at least from the process $\pi^0
\rightarrow \gamma\gamma$. Then $-Im \Pi^R \simeq \Gamma
(\gamma\gamma \rightarrow \pi^0 )\sim \Gamma^{vac} (\gamma\gamma
\rightarrow \pi^0 )$. Although more involved effects may also
simulate the corresponding width terms, one may expect that the
width effects are, nevertheless, rather suppressed due to the
nature of the superconducting phase with the large gap.  In this
situation the pion pole mechanism could become an efficient
cooling mechanism. Also other meson poles may essentially
contribute.
%%In order to
%%demonstrate the effect we took
%
%%\be \mid D_{\pi^0}^R\mid^2=\frac{1}{(\omega^2 -k^2 /3
%%-m_{\pi^0}^{*2} )^2 + (Im \Pi^R )^2 } \label{scprop} \ee
%
%%and calculated the emissivity of the CFL phase of the
%%superconducting medium $Q^{SC}$.
%% for temperatures in the interval
%%$0.01\div 50$~MeV with $m^*_{\pi}=50$~MeV and $m^*_{\pi}=10$~MeV
%%taking $-Im \Pi^R /m^*_{\pi}\simeq 0.01$~MeV, $-Im \Pi^R
%%/m^*_{\pi}\simeq 0.1$~MeV
%% and $-Im \Pi^R /m^*_{\pi}
%%\simeq 1$~MeV, assuming that there are still processes related to
%%the strong interaction which although being rather suppressed
%%nevertheless contribute to the width. If there were no such
%%processes, one should take $-Im \Pi^R \simeq \Gamma (\gamma\gamma
%%\rightarrow \pi^0 )$.

%
%%\begin{figure}[ht]
%%\begin{center}
%%\includegraphics{figura2.eps}
%%\caption[dummy0]{Ratio of energy loss $Q^{SC}/Q^{vac}$ as function
%%of temperature for $-Im \Pi^R /m^*_{\pi} \simeq 1$~MeV (dashed
%%line), $-Im \Pi^R /m^*_{\pi} \simeq 0.1$~MeV (long dashed line)
%%and $-Im \Pi^R /m^*_{\pi} \simeq 0.01$~MeV (dot-dashed line) for
%%$m^*_{\pi}=10$~MeV.}
%%\end{center}
%%\label{fig2}
%%\end{figure}
%
%
%%\begin{figure}[ht]
%%\begin{center}
%%\includegraphics{figura3.eps}
%%\caption[dummy0]{Ratio of energy loss $Q^{SC}/Q^{vac}$ as function
%%of temperature for $-Im \Pi^R /m^*_{\pi} \simeq 1$~MeV (dashed
%%line), $-Im \Pi^R /m^*_{\pi} \simeq 0.1$~MeV (long dashed line)
%%and $-Im \Pi^R /m^*_{\pi} \simeq 0.01$~MeV (dot-dashed line) for
%%$m^*_{\pi}=50$~MeV.}
%%\end{center}
%%\label{fig3}
%%\end{figure}
%

\begin{figure}[ht]
\begin{center}
\includegraphics{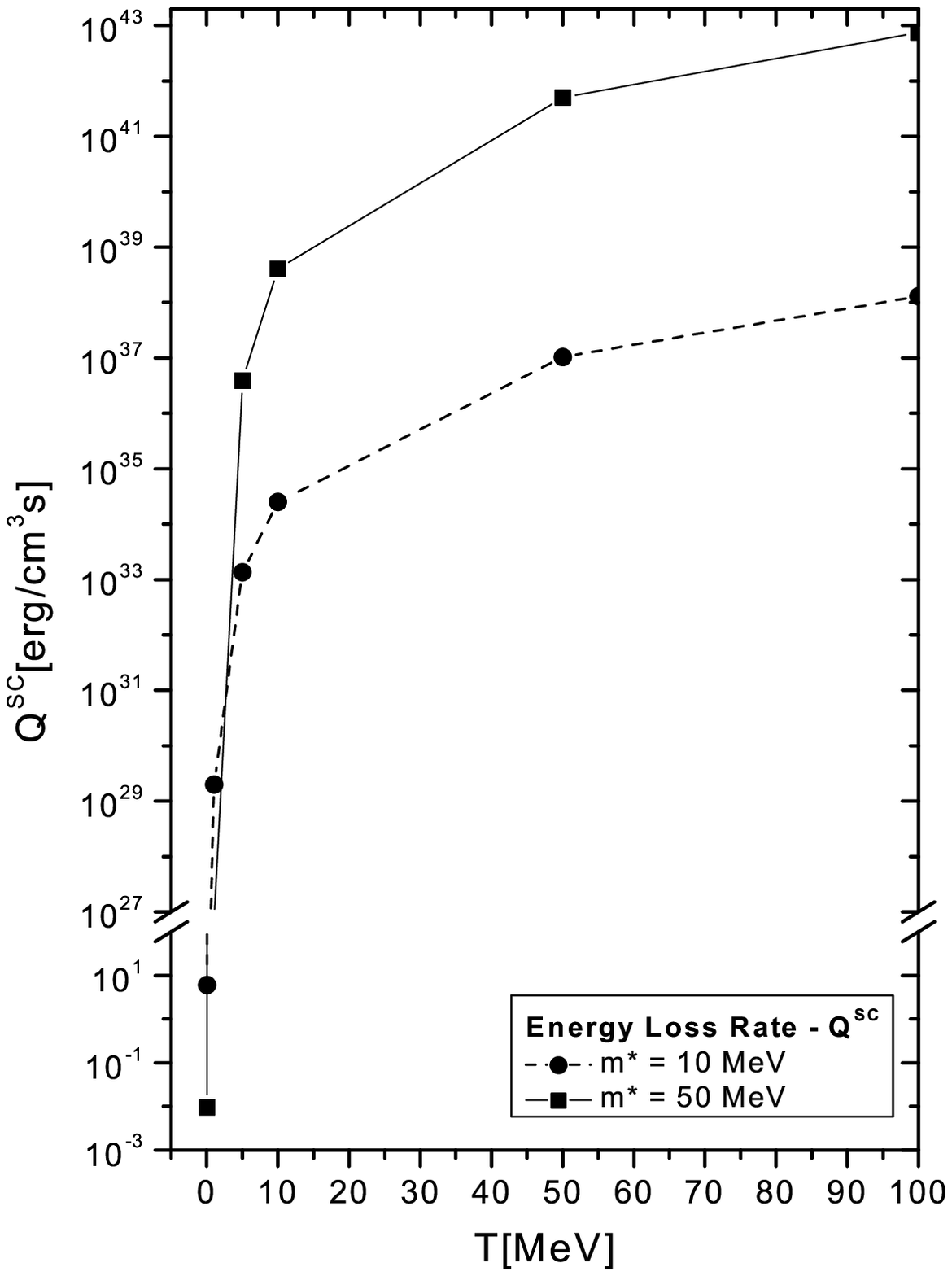}
\caption[dummy0]{Emissivity $Q^{SC}$ in $erg/(cm^{3}\cdot s)$
computed with the interpolation formula of Eq.(\ref{est}) for
$m^*_{\pi}=10$ and $50$~MeV. The curves were obtained assuming the
total width $\Gamma_\pi \simeq \Gamma^{vac}_\pi$ . } \label{fig2}
\end{center}
\end{figure}

For the process under consideration the emissivity of the
superconducting medium $Q^{SC}$ is given by expressions similar to
Eqs. (\ref{qrate}) to (\ref{qpi}). We take
\be \mid D_{\pi^0}^R\mid^2=\frac{1}{(\omega^2 -k^2 /3
-m_{\pi^0}^{*2} )^2 + (Im \Pi^R )^2 } \label{scprop} \ee
and replace it
%(\ref{scprop})
into (\ref{crosssec}) instead of (\ref{piprop}). Other quantities
in (\ref{crosssec}) are assumed to be unchanged. Then the energy
loss rate  is presented as
\br Q^{SC} \simeq 4.2 \times 10^{22}
%\(\frac{16}{\pi^3}\)
\(\frac{\Gamma (\pi^0 \rightarrow \gamma \gamma)}{m_{\pi}}\)
\(\frac{\Gamma (\pi^0 \rightarrow \nu \bar{\nu})}{m_{\pi}}\)
%\(\frac{1 \, GeV^2}{m_{\pi}^2}\)\\
%\nonumber && \times
%\( \frac{k_B T}{m_{\pi}}  \)^4  (k_B T)^7
T_9^{11}(\frac{m_{\pi}}{m_{\pi}^{*}})^4 I^{SC}(\widetilde{\tau}
),\,\,\, \frac{erg}{cm^3 \cdot s},  \l{qpiSC} \er
now with the extra factor $(\frac{m_{\pi}}{m_{\pi}^{*}})^4$ and
the integral $I(\tau)$ replaced by
\br \l{integ2} I^{SC}(\widetilde{\tau} ) &=& \int_0^\infty
\frac{x_1^4 \, dx_1}{e^{x_1} - 1}\int_0^\infty \frac{x_2^4 \,
dx_2}{e^{x_2} - 1}\nonumber
\\ &\times& \int_\kappa^2 \frac{du(x_1+x_2)
u^3}{(\frac{2}{3}\widetilde{\tau}^2[ (x_{1}+x_{2})^{2}
+x_{1}x_{2}u] - 1)^2 + (-m_{\pi}^{*-2}Im\Pi^R_{\pi^0})^2}  ,   \er
$\widetilde{\tau}=T/m_{\pi}^*$.  Again we further take $\kappa=0$.
For $\widetilde{\tau} \ll 0.1 $ we get \br\label{asymSC}
I^{SC}(\tau\ll 0.1)\simeq 23040 \,\zeta (6)\zeta (5)\left[
1+92\tau^2 \left(\frac{7\zeta(8)}{23\zeta(6)} +\frac{\zeta
(7)}{\zeta (5)}\right)\right]\,,\er that not essentially differs
from the estimation given by Eq.(\ref{asym}), $\zeta(8)=1.004$.
The value (\ref{asymSC}) is independent on $m_{\pi}^*$ for $\tau
\ll 0.1$, and $Q^{SC}\propto T^{11}(m_{\pi}^*)^{-4}$,
$Q^{SC}/Q^{vac}\sim (m_\pi /m^{*}_{\pi})^4$, cf. (\ref{vacq}). For
$[m^{*\,2}_{\pi}/(-10^2 Im \Pi^R )]^{1/4}\gsim
\widetilde{\tau}\gsim 1/a$ we roughly estimate $I^{SC}\sim 0.1
[-m_{\pi}^{*-2} \mbox{Im}\Pi^R ]^{-1}
\widetilde{\tau}^{-8}(e^{-1/(2\widetilde{\tau})}
|\mbox{ln}(2\widetilde{\tau} )|+1)e^{-1/(2\widetilde{\tau})}$,
$a\sim 10^2$ for $-\mbox{Im}\Pi^R \sim \Gamma^{vac}$, and
$Q^{SC}\propto T^{3}(m_{\pi}^*)^{6}e^{-m^*/(2T)}/(-\mbox{Im}\Pi^R
)$. For still higher temperatures $I^{SC}\simeq 8
/\widetilde{\tau}^4$ and $Q^{SC}\propto T^{7}$ being almost
independent on $m^*_{\pi}$, $Q^{SC}\simeq 1.3\cdot
10^{29}\left(m_{\pi}^{-1}\Gamma( {\pi}^0 \rightarrow
\nu\bar{\nu})\right)T_9^{7},\,\,erg/(cm^3 \cdot s)$. Interpolation
estimation being roughly valid for all temperatures produces \br
\label{est} I^{SC}\simeq \frac{24299(1+120\widetilde{\tau}^2
e^{-10\widetilde{\tau}})}{1+2880\widetilde{\tau}^4} +\frac{0.1
(e^{-1/(2\widetilde{\tau})} |\mbox{ln}(2\widetilde{\tau}
)|+1)e^{-1/(2\widetilde{\tau})}}{\widetilde{\tau}^{8}[-m_{\pi}^{*-2}
\mbox{Im}\Pi^R ]}.
 \er
The emissivity  ($Q^{SC}$) computed with the above interpolation
formula (Eq.(\ref{est})) and for the total pion width given by the
vacuum one is shown in Fig.\ref{fig2}. The value $\Gamma (\pi^0
\rightarrow \nu \bar{\nu})$ is assumed to be equal to its
experimental upper limit. This figure serves as a guide for the
numerical calculation of Eq.(\ref{integ2}), showing the very fast
increase of the emissivity as we approach temperatures near the
pion pole one and demonstrating a slow increase as we go to
temperatures above the effective pion mass scale. The
interpolation formula and the full numerical calculation of the
emissivity agree with high accuracy for small $T$ (compared to
$m^*_{\pi}$) and differ by a factor $\lsim 2$ for $T \approx
m^*_{\pi}$.

\begin{figure}[ht]
\begin{center}
\includegraphics{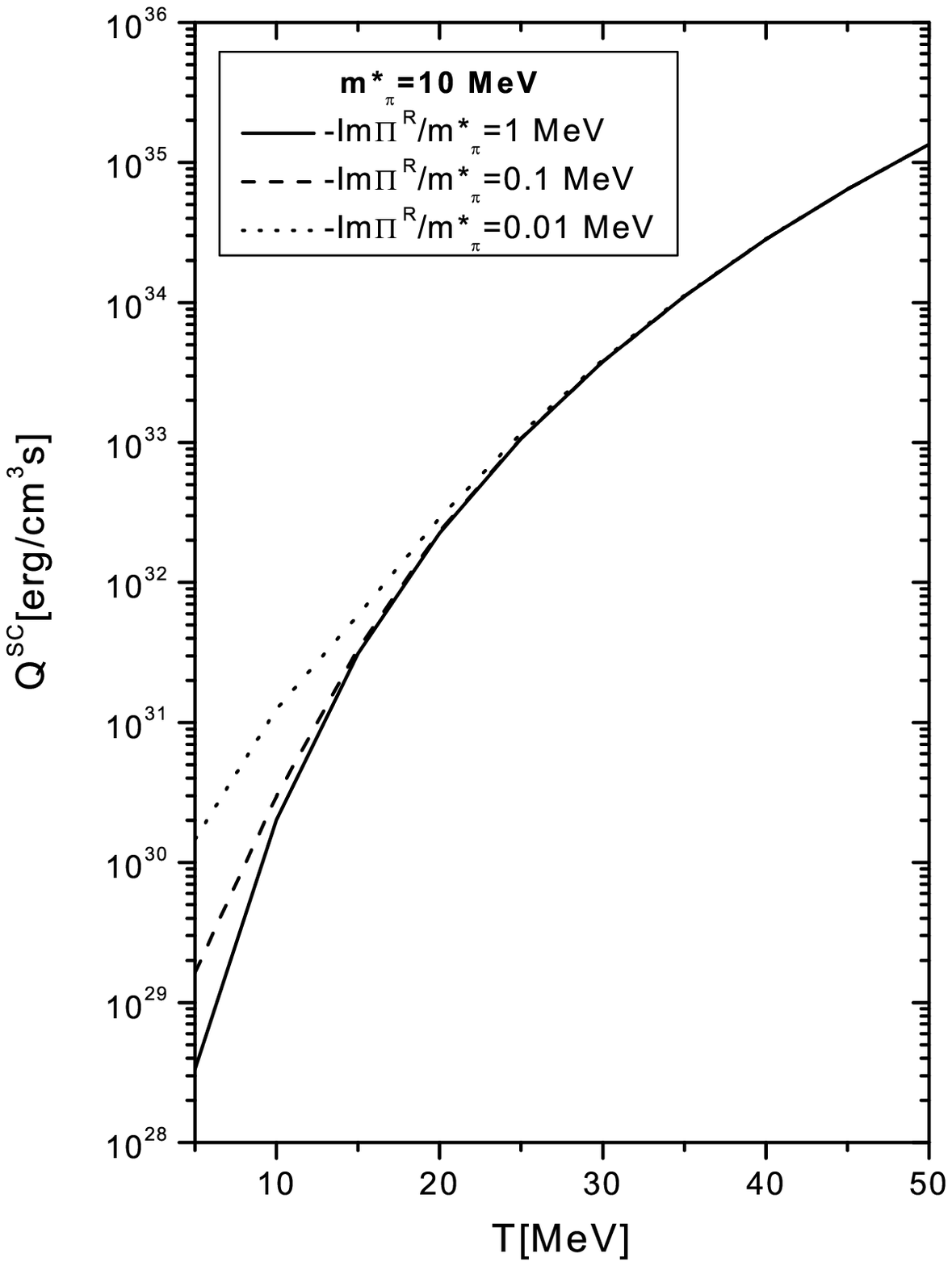}
\caption[dummy0]{Emissivity $Q^{SC}$ in $erg/(cm^{3}\cdot s)$ as
function of temperature for $-Im \Pi^R /m^*_{\pi} \simeq 1$~MeV
(solid line), $-Im \Pi^R /m^*_{\pi} \simeq 0.1$~MeV (dashed line)
and $-Im \Pi^R /m^*_{\pi} \simeq 0.01$~MeV (dotted line) for
$m^*_{\pi}=10$~MeV.} \label{fig3}
\end{center}
\end{figure}

\begin{figure}[ht]
\begin{center}
\includegraphics{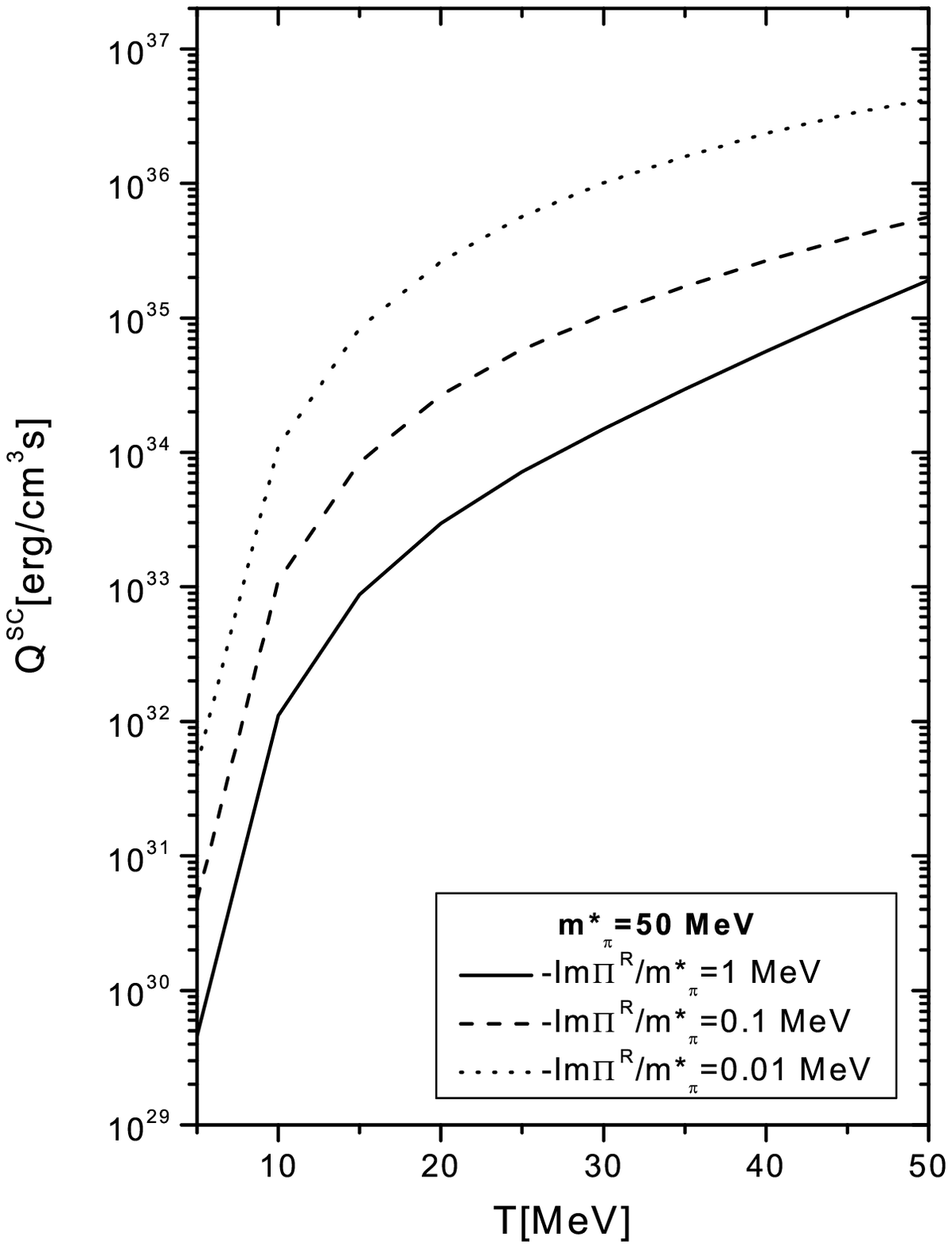}
\caption[dummy0]{The same as in Fig. \ref{fig4} but for $m^*_{\pi}
=50$~MeV.} \label{fig4}
\end{center}
\end{figure}

In Figs. \ref{fig3}, \ref{fig4} we show the numerically calculated
emissivity of the process under consideration for the
superconducting media, $Q^{SC}$, in the temperature range from 5
to 50 MeV for $m^*_{\pi}=10$ MeV and $m^*_{\pi}=50$ MeV,
respectively. These curves were obtained assuming $\Gamma (\pi^0
\rightarrow \nu \bar{\nu})$ equal to its experimental upper limit.
In the given temperature interval (from 5 to 50 MeV ) the
emissivity curves are scaled up for $m^*_{\pi}=50$ MeV compared to
$m^*_{\pi}=10$ MeV in accordance with above estimations of
$I^{SC}$ and interpolation equation (\ref{est}). The pole
asymptotic manifests itself even in the case when the width is rather
suppressed (solid curves). It happens for $T< (15\div 20)$~MeV in
case $m^*_{\pi}=10$ MeV and in the whole temperature interval (
from 5 to 50 MeV) in case $m^*_{\pi}=50$ MeV. For  $m^*_{\pi}=10$
MeV solid, dash and dotted curves reach the high temperature
asymptotic for $T>20$~MeV. For $m^*_{\pi}=50$ MeV the solid curve
achieves the high temperature asymptotic for $T>50$~MeV. The
smaller is the width the more pronounced is the pole mechanism.  This
is clearly seen if we compare Fig.\ref{fig2} (the result of the
interpolation formula computed for the vacuum width) with  Figs.
\ref{fig3}, \ref{fig4} where we assumed that more involved effects
may simulate a width larger than the vacuum one.
%down by at least three orders of magnitude.
%%Note that as long as $-Im \Pi^R
%%/m^*_{\pi} << m^*_{\pi}$ the width effect is not important for the
%%emissivity, and the efficiency of the pion pole mechanism depends
%%mainly on the ratio between the temperature and the pion mass.

%
%

%The reason for the enhancement is not difficult to understand.
Compared to the hadron case, in the color-flavor-locked phase of
the superconducting quark medium  the effective pion mass is much
smaller  and the width is suppressed. In this case the resonance
effect appears with a larger amplitude (due to very small width)
and at smaller temperatures. Thus we argue that for the star core
in the color-flavor-locked phase at $T<T_{opac}^{SC}$, i.e. beyond
the neutrino opacity regime, in absence of any other  efficient
cooling reaction channels the pion pole mechanism could become an
efficient cooling mechanism even if $\Gamma (\pi^0 \rightarrow
\nu\bar{\nu})$ being by many orders of magnitude below the value
determined by the modern experimental upper limit. Here the value
of $T_{opac}^{SC}$ is different from that for the usual
proto-neutron star being determined by the condition that the
neutrino mean free path in the quark core becomes to be comparable
with the size of the core. In the neutrino opacity regime,
$T>T_{opac}^{SC}$, the reaction $\pi^0 \rightarrow \nu\bar{\nu}$
delays the neutrino transport from the quark core to the hadron
shell. To come to these conclusions we used only that the
effective pion mass in the CFL phase is essentially smaller than
the vacuum one, the pion width is small and the pion momentum is
shifted by $k\rightarrow k/\sqrt{3}$. In some relevant temperature
interval for the width $\Gamma_\pi \sim \Gamma^{vac}_\pi$ and for
not too small effective pion mass  the rate $Q^{SC}$
%for $m^*_{\pi}\lsim 10$~MeV and for the width $-Im \Pi^R
%/m^*_{\pi}\lsim (m^*_{\pi}/10)$~MeV the rate $Q^{SC}$
%(which grows as $(T/m_{\pi})^{11}$)
may even   exceed the emissivity of the most efficient direct Urca
process being $Q^{DU}\sim 10^{39}(T/10~\mbox{MeV})^6$~ $erg/(cm^3
\cdot s)$ for the non-superfluid (hadron) neutron star matter.
Also at very low temperatures the emissivity of the process
 decreases with the temperature according to
the power low ($Q^{SC}\propto T^{11}$) rather than exponentially
as other relevant processes in the CFL phase. Therefore in spite
of a low absolute value of the rate at such low temperatures, the
process is the dominating process.

%%We also verified the efficiency of the mechanism in the low
%%temperature superconducting regime computing the emissivities for
%%temperatures in the range $T \approx 0.01-0.1$~MeV for $m^*_{\pi}$
%%equal to 10 and 50 MeV. We found that in this limit the emissivity
%%does not depend strongly on the value of   $-Im \Pi^R /m^*_{\pi}$
%%and a reasonable estimate for the output of energy can be obtained
%%observing that
%
%%\be \frac{Q^{SC}}{Q^{vac}} \propto \left(
%%\frac{m_\pi}{m^*_\pi}\right)^4 ,
%% \l{ratiolow}
%% \ee
%
%%see (\ref{qpiSC}), (\ref{integ2}), (\ref{asym}). Therefore the
%%enhancement of $Q^{SC}$ compared
%% to  $Q^{vac}$ at one given low
%%temperature will basically depend on the value of $m^*_\pi$. The
%%smaller is $m^*_\pi$ the more efficient will be the mechanism.

Now we may compare the emissivity of the process
$\gamma\gamma\rightarrow \pi^0\rightarrow \nu\bar\nu$
%%given by Eq.
%%(\ref{qpi}) with $I(\tau)$ from (\ref{integ2}),
with the estimation for the emissivity (\ref{pin}) of the process
$\pi^0\rightarrow \nu\bar\nu$ in the quark matter discussed in
\cite{BKV,PR,RS}. The emissivity of the later process is given by
Eq.(\ref{pin})
%
%% \br\label{pin} Q^{\pi^0\nu\bar\nu}\sim \left( \frac{\Gamma
%% (\pi^0\rightarrow \nu\bar\nu )}{m_{\pi}} \right)
%% n_{\pi}m_{\pi}^{*2} , \er
%% $$n_{\pi}=\frac{1}{v_{\pi}^3}\left(\frac{
%% m^*_{\pi}k_B T}{2\pi}\right)^{3/2}e^{-m^*_{\pi}/(k_B T)},$$ where
with $v^2_{\pi}=1/3$ for the color-flavor-locked phase. We see
that the ratio (\ref{qpiSC}) to (\ref{pin}) is larger than unit in
a wide temperature interval of our interest, $k_B T\lsim 7$~MeV
for $m^*_{\pi}\sim 70$~MeV and $k_B T\lsim 0.2$~MeV for
$m^*_{\pi}\sim 10$~MeV. In this estimation we again used that the
total pion width is almost exhausted by the $\gamma\gamma$ decay,
$\Gamma (\gamma\gamma \rightarrow \pi^0 ) \simeq -Im \Pi^R$ and,
as before, we took $\Gamma (\gamma\gamma \rightarrow \pi^0 )\simeq
\Gamma_{\pi}^{vac}$. If the width $\Gamma_{\pi}$ were larger as
the consequence of the presence of some  processes which were not
considered up to now, then the pole contribution could be more
suppressed (see dotted, dash and solid curves in Figs \ref{fig3},
\ref{fig4}) and the process $\gamma\gamma\rightarrow
\pi^0\rightarrow \nu\bar\nu$ would be
%more effective compared to the process $\pi^0\rightarrow
%\nu\bar\nu$
relevant only for low temperatures $T\lsim (0.1\div 1)$~MeV.

Notice that the ratio of the reaction rates for
$\gamma\gamma\rightarrow \pi^0\rightarrow \nu\bar\nu$ and
$\pi^0\rightarrow \nu\bar\nu$ does not depend on the value of
$\Gamma (\pi^0\rightarrow \nu\bar\nu)$ and it always becomes
larger than unit with the decrease of  the temperature (however
the critical temperature when the rate reaches the unit depends on
the values of the parameters).

Refs \cite{PR,RS} found an interesting possibility that in the
color-flavor-locked phase the process $\pi^0\rightarrow
\nu\bar\nu$ is allowed also in the standard model with left-handed
neutrinos even if the neutrino mass is zero. The important point
that was noticed is that the temporal and thermal components of
the pion decay constant need not be the same in the dense medium
due to the violation of the Lorentz invariance. The calculation of
\cite{PR} yields
\br\label{pin1} \Gamma (\pi^0\rightarrow \nu\bar\nu
)&=&\frac{1}{4\pi v_{\pi}^2}G_F^2 (f_T -f_S )^2 k_B T\,
m_{\pi}^{*} m_{\pi}\nonumber \\ &\simeq& 0.6\cdot
10^{-20}\left(\frac{\mu_q}{10~MeV}\right)^2  \frac{m^*_{\pi}}{10
MeV }\left(\frac{k_B T}{10~MeV}\right)m_{\pi} ,\er $\mu_q$ is the
quark chemical potential. We see that for $m^*_{\pi}\sim k_B T\sim
10$~MeV the value (\ref{pin1}) is by three-four orders of
magnitude smaller than the value of the experimental upper limit
for $\Gamma (\pi^0\rightarrow \nu\bar\nu)$ for the right-handed
neutrinos. Therefore both possibilities should be studied.

Ref. \cite{PR} also evaluated the value $\Gamma (\gamma\gamma
\rightarrow \pi^0 )$ for the corresponding decay in the
color-flavor-locked phase. This estimation only by the factor
$\sim 1$ differs from the value $\Gamma (\gamma\gamma \rightarrow
\pi^0 )\simeq \Gamma_{\pi}^{vac}$ which we used in our
estimations.

\section{Conclusion}
In conclusion, (i) we evaluated the cooling rate of the
proto-neutron stars  via the pion pole mechanism taking into
account the nuclear medium effects and we showed that with the
width effects included this mechanism gives no constraints on the
corresponding $\pi^0 \rightarrow \nu_R\bar{\nu}_L (\nu_L
\bar{\nu}_R)$ decay width. (ii) We also discussed possible
contribution of this mechanism to the cooling of different
astrophysical systems, as is the case of neutron stars with dense
quark cores. We found that  the process $\gamma\gamma\rightarrow
\pi^0\rightarrow \nu\bar\nu$ is proved to be the most
 efficient process in the color-flavor-locked
superconducting core at temperatures $T\lsim (0.1\div 10)$~MeV
depending on the effective pion mass and the pion decay width.
Depending on what the future tells us about $\Gamma (\pi^0
\rightarrow \nu \bar{\nu})$ and on the possibility of the color
superconductivity in neutron stars we believe that these questions
deserve a further more detailed study (perhaps even further
studying of a dependence on medium effects of all the terms
appearing in the cross section of $\gamma\gamma \rightarrow \pi^0
\rightarrow \nu_R\bar{\nu}_L (\nu_L \bar{\nu}_R)$).

\section*{Acknowledgments}

We would like to thank A. C. Aguilar for the help in the
numerical procedures and E.E. Kolomeitsev for valuable remarks.
This research was supported in part by the Conselho Nacional de
Desenvolvimento Cient\'{\i}fico e Tecnol\'ogico (CNPq) (AAN), and
in part by Coordenadoria de Aperfei\c coamento do Pessoal de
n\'{\i}vel Superior (FA).

\appendix
\section{Optical theorem formalism}
In  \cite{VS87} and then in \cite{KV95}, see \cite{dv} for a
review, there was developed the optical theorem formalism in terms
of full non-equilibrium Green functions to calculate the reaction
rates including finite particle widths and other in-medium
effects. Applying this approach, e.g., to the
antineutrino--neutrino production we can express the transition
probability in a direct reaction in terms of the evolution
operator $S$,
\begin{equation}\label{optfirst}
\frac{d{\cal W}^{\rm  tot}_{X\rightarrow \bar\nu \nu}}{d t}=
\frac{d q_{\nu}^3 dq_{\bar{\nu}}^3}{(2\pi)^6\,4\, \omega_{\nu}
\,\omega_{\bar{\nu}}} \,\sum_{\{X\}}\overline{<0|\, S^\dagger\,
|\bar\nu \nu + X>\,<\bar\nu \nu +X|\, S\, |0>}\,,
\end{equation}
where we presented explicitly the phase-space volume of $\bar\nu
\nu$ states; lepton occupations of given spin are put equal to
zero for $\nu$ and $\bar{\nu}$ which are supposed to be radiated
directly from the system. The bar denotes statistical averaging.
The summation goes over the complete set of all possible
intermediate states $\{X\}$ constrained by the energy-momentum
conservation. Making use of the smallness of the weak coupling, we
expand the evolution operator as
%\begin{equation}\label{Smatrix}
$S\approx 1- i \, \intop_{-\infty}^{+\infty} T\,\bigl\{V_W(x)\,
S_{\rm nucl} (x)\bigr\} d x_0 \,,$
%\end{equation}
where $V_W$ is the Hamiltonian of the weak interaction
%taken in
%Eq.~(\ref{Smatrix})
in the interaction representation, $S_{\rm nucl}$ is the part of
the $S$ matrix corresponding to the nuclear interaction, and
$T\{...\}$
%stands for
is the chronological  ordering operator. After substitution into
(\ref{optfirst}) and averaging over the arbitrary non-equilibrium
state of a nuclear system, there appear chronologically ordered
($G^{--}$), anti-chronologically ordered ($G^{++}$) and disordered
($G^{+-}$ and $G^{-+}$) exact Green functions.

In  graphical form the general expression for the probability of
the  neutrino and anti-neutrino production is as follows
\\
\begin{eqnarray}\label{sm}
\parbox{10mm}{\setlength{\unitlength}{1mm}
\begin{fmfgraph*}(30,10)
\fmfleftn{l}{2} \fmfrightn{r}{2}\fmfpen{thick}
\fmfpoly{hatched,pull=1.4,smooth}{ord,oru,olu,old}
\fmfforce{(0.3w,.8h)}{olu} \fmfforce{(0.7w,.8h)}{oru}
\fmfforce{(0.3w,0.2h)}{old} \fmfforce{(0.7w,0.2h)}{ord}
\fmfforce{(0.25w,0.5h)}{ol} \fmfforce{(0.75w,0.5h)}{or}
\fmf{fermion,width=1thin}{l1,ol} \fmf{fermion,width=1thin}{ol,l2}
\fmf{fermion,width=1thin}{r2,or} \fmf{fermion,width=1thin}{or,r1}
\fmflabel{$\bar\nu$}{l1} \fmflabel{$\nu$}{l2}
\fmflabel{$\bar\nu$}{r2} \fmflabel{$\nu$}{r1}
\fmfv{l=$+$,l.a=180,l.d=3thick}{ol}
\fmfv{l=$-$,l.a=0,l.d=3thick}{or}
\end{fmfgraph*}}
\,\,\, ,
\end{eqnarray}
%\noindent
\\
representing the sum of all closed diagrams ($-i \Pi^{-+}$)
containing at least one ($-+$) exact Green function. The latter
quantity is especially important. Various contributions from
$\{X\}$ can be classified according to the number $N$ of $G^{-+}$
lines in the diagram
\\
\begin{eqnarray}\label{qp}
%\nonumber
\frac{d{\cal W}^{\rm  tot}_{\bar\nu \nu}}{d t}= \frac{d^3
q_{\bar{\nu}} d q_{\nu}^{3} }{(2\pi
)^6\,4\,\omega_{\bar{\nu}}\,\omega_{\nu}} \left(\,\,
\parbox{27mm}{
\setlength{\unitlength}{1mm}
\begin{fmfgraph*}(27,10)
\fmfleftn{l}{2} \fmfrightn{r}{2}\fmfpen{thick}
\fmfpoly{empty,pull=1.4,smooth,label= $N=1$ }{ord,oru,olu,old}
\fmfforce{(0.3w,.8h)}{olu} \fmfforce{(0.7w,.8h)}{oru}
\fmfforce{(0.3w,0.2h)}{old} \fmfforce{(0.7w,0.2h)}{ord}
\fmfforce{(0.25w,0.5h)}{ol} \fmfforce{(0.75w,0.5h)}{or}
\fmf{fermion,width=1thin}{l1,ol} \fmf{fermion,width=1thin}{ol,l2}
\fmf{fermion,width=1thin}{r2,or} \fmf{fermion,width=1thin}{or,r1}
\fmflabel{$\bar\nu$}{l1} \fmflabel{$\nu$}{l2}
\fmflabel{$\bar\nu$}{r2} \fmflabel{$\nu$}{r1}
\fmfv{l=$+$,l.a=180,l.d=3thick}{ol}
\fmfv{l=$-$,l.a=0,l.d=3thick}{or}
\end{fmfgraph*}
} \,+\,
\parbox{27mm}{
\setlength{\unitlength}{1mm}
\begin{fmfgraph*}(27,10)
\fmfleftn{l}{2} \fmfrightn{r}{2}\fmfpen{thick}
\fmfpoly{empty,pull=1.4,smooth,label= $N= 2$ }{ord,oru,olu,old}
\fmfforce{(0.3w,.8h)}{olu} \fmfforce{(0.7w,.8h)}{oru}
\fmfforce{(0.3w,0.2h)}{old} \fmfforce{(0.7w,0.2h)}{ord}
\fmfforce{(0.25w,0.5h)}{ol} \fmfforce{(0.75w,0.5h)}{or}
\fmf{fermion,width=1thin}{l1,ol} \fmf{fermion,width=1thin}{ol,l2}
\fmf{fermion,width=1thin}{r2,or} \fmf{fermion,width=1thin}{or,r1}
\fmflabel{$\bar\nu$}{l1} \fmflabel{$\nu$}{l2}
\fmflabel{$\bar\nu$}{r2} \fmflabel{$\nu$}{r1}
\fmfv{l=$+$,l.a=180,l.d=3thick}{ol}
\fmfv{l=$-$,l.a=0,l.d=3thick}{or}
\end{fmfgraph*}
} \dots\right)\,\, .
\end{eqnarray}
%\noindent
\\
Being expressed in terms of the exact Green functions, each
diagram in (\ref{qp}) represents a whole class of perturbative
diagrams of any order in the interaction strength and in the
number of loops. This procedure suggested in \cite{VS87} is
actually very helpful especially if the quasiparticle
approximation holds for the intermediate fermions or bosons. Then
contributions of specific processes contained in a closed diagram
can be made visible by cutting the diagrams over the ($+-$),
($-+$) lines. In the framework of the quasiparticle approximation
for the non-relativistic fermions $G^{-+}_{F}=2\pi i n_{F}\delta
(\varepsilon +\mu -\varepsilon^0_p -\mbox{Re}\Sigma^R (\varepsilon
+\mu , \vec{p}))$ ($n_F$ are fermionic occupations, for
equilibrium $n_F = 1/[\mbox{exp} ((\varepsilon -\mu_F )/T)+1]$),
and the cut eliminating the energy integral thus requires clear
physical meaning. In this way one establishes the correspondence
between closed diagrams and usual Feynman amplitudes although in
the general case of finite fermion width the cut has only a
symbolic meaning. Next advantage is that in the quasiparticle
approximation any extra $G^{-+}_{F}$, since it is proportional to
$n_F$, brings a small $(T/\varepsilon_F )^2$ factor to the
emissivity of the process. Dealing with small temperatures one can
restrict by the diagrams of the lowest order in
$(G^{-+}_{F}G^{+-}_{F})$, not forbidden by energy-momentum
conservations, putting $T=0$ in all $G^{++}$ and $G^{- -}$ Green
functions. For the relativistic bosons with the non-conserving
number, in the quasiparticle approximation $G^{-+}_{B}=-2\pi i
n_{B}\delta \left(\omega^2 -m_B^2 -k^2 -\mbox{Re}\Pi^R (\omega , k
)\right)$ ($n_B$ are bosonic occupations, for equilibrium $n_B =
1/[\mbox{exp} (\omega/T)-1]$). In a wide temperature and
energy-momentum region $\mbox{Im}\Pi^R$ is not small and the
quasiparticle approximation is then not valid for the boson Green
functions whereas the region, where $\mbox{Im}\Sigma^R $ is small
and the quasiparticle approximation is valid for the fermion Green
functions, is usually much wider.

If one is interested only in  the processes related to the
$\nu\bar{\nu}$ coupling with the $\pi^0$, the hatched block in
(\ref{sm}) is reduced to the exact $D^{-+}_{\pi^0}$ Green
function. This Green function satisfies the exact Dyson equation.
Only if in a specific region of the pion energies and momenta
($\omega$ and $k$) the pion Green function $D^{-+}_{\pi^0}$ can be
approximated by the $\delta$-function, integrating over this
region one may use the quasiparticle approximation for the pion.
In such a way one usually calculates the rate of the $\pi^0
\rightarrow \nu\bar{\nu}$ process. Certainly, in other regions of
pion energies and momenta the pion Green function contains the
width relating to different channels of the pion decay. E.g. the
polarization operator of the $\pi^0$ contains the $"-+"$
$\gamma\gamma$ loop. This term corresponds to the contribution
$D^{--}_{\pi^0}G_{\gamma}^{-+}G_{\gamma}^{+-}D^{++}_{\pi^0}$.
Within the quasiparticle approximation to the $\gamma$ one cuts
the $G_{\gamma}^{-+}G_{\gamma}^{+-}$ lines and gets in this way
the contribution of the $\gamma\gamma \rightarrow \pi^0
\rightarrow \nu\bar\nu$ process which we discuss in this paper.
When considering right-handed neutrinos we do not know the
explicit expression for the $\pi^0 \rightarrow \nu\bar\nu$ vertex.
Several different expressions can be used. Therefore we will
express the result of the integration in the $\nu\bar{\nu}$ states
in (\ref{optfirst}) via the phenomenological value of the width
$\Gamma (\pi^0 \rightarrow \nu\bar{\nu})$  for which there exists
the experimental upper limit. If we knew the coupling we could
present an explicit calculation, as one usually does for the
left-handed neutrinos.

\end{fmffile}
%\newpage

%

\end{document}